\documentclass[10pt,conference]{IEEEtran} 
\IEEEoverridecommandlockouts
\usepackage{cite}
\usepackage{nicefrac}
\usepackage{amsmath,amssymb,amsfonts}
\usepackage{physics}
\usepackage{algorithmic}
\usepackage{graphicx}
\graphicspath{{./figures/}}
\usepackage{multirow}
\ifCLASSOPTIONcompsoc \usepackage[caption=false,font=normalsize,labelfon
t=sf,textfont=sf]{subfig}
\else
\usepackage[caption=false,font=footnotesize]{subfi g}
\fi
\usepackage{textcomp}
\usepackage{xcolor}
\def\BibTeX{{\rm B\kern-.05em{\sc i\kern-.025em b}\kern-.08em
    T\kern-.1667em\lower.7ex\hbox{E}\kern-.125emX}}
\begin{document}

\title{Time-Efficient Qudit Gates through \\ Incremental Pulse Re-seeding
\thanks{This work is funded in part by EPiQC, an NSF Expedition in Computing, under award CCF-1730449; in part by STAQ under award NSF Phy-1818914; in part by NSF award 2110860; in part by the US Department of Energy Office 
of Advanced Scientific Computing Research, Accelerated Research for Quantum Computing Program; and in part by the NSF Quantum Leap Challenge Institute for Hybrid Quantum Architectures and Networks (NSF Award 2016136) and in part based upon work supported by the U.S. Department of Energy, Office of Science, National Quantum Information Science Research Centers.  

FTC is Chief Scientist for Quantum Software at ColdQuanta and an advisor to Quantum Circuits, Inc.

Correspondence: lmseifert@uchicago.edu, jchadwick@uchicago.edu}
}

\author{\IEEEauthorblockN{Lennart Maximilian Seifert\IEEEauthorrefmark{2}\IEEEauthorrefmark{1}, Jason Chadwick\IEEEauthorrefmark{3}\IEEEauthorrefmark{1}, Andrew Litteken\IEEEauthorrefmark{2}, Frederic T. Chong\IEEEauthorrefmark{2} and Jonathan M. Baker\IEEEauthorrefmark{2}} \IEEEauthorblockA{\IEEEauthorrefmark{2}Department of Computer Science, University of Chicago} \IEEEauthorblockA{\IEEEauthorrefmark{3}Department of Physics, Carnegie Mellon University}
\IEEEauthorblockA{\IEEEauthorrefmark{1}These authors contributed equally.}
}

\maketitle

\begin{abstract}


Current efforts to build quantum computers focus mainly on the two-state qubit, which often involves suppressing readily-available higher states. In this work, we break this abstraction and synthesize short-duration control pulses for gates on  generalized \textit{d}-state qu\textit{dits}. We present Incremental Pulse Re-seeding, a practical scheme to guide optimal control software to the lowest-duration pulse by iteratively seeding the optimizer with previous results. We find a near-linear relationship between Hilbert space dimension and gate duration through explicit pulse optimization for one- and two-qudit gates on transmons. Our results suggest that qudit operations are much more efficient than previously expected in the practical regime of interest and have the potential to significantly increase the computational power of current hardware.

\end{abstract}




\section{Introduction}
Quantum computing traditionally focuses on the realization of noise-robust two-level systems, known as qubits. However, in many quantum architectures, each qubit is embedded in a much larger Hilbert space, with all other energy levels being ignored or suppressed. 
Qudits, the extension of qubits to $d$ levels, are a promising topic of study with the potential to increase computational power of a machine without needing to add additional logic units. For instance, a single four-state qudit can store the same amount of information as two qubits, and an eight-state qudit can encode the global state of three qubits. There are many proposed or adapted qudit-based quantum algorithms \cite{gedik_computational_2015, deller_quantum_2022, nguyen_quantum_2019, nagata_generalization_2020, bocharov_factoring_2017} and qudit-based improvements in quantum circuit compilation \cite{gokhale_asymptotic_2019, baker_efficient_2020}. The latter presents a method of using qudits in intermediate steps of a circuit that asymptotically reduces the required number of ancilla in a quantum algorithm at essentially equal circuit depth.

Qudits have been studied in different experimental settings. Both IBM \cite{galda_implementing_2021} and Rigetti  \cite{rigetti_computing_beyond_2021} have demonstrated implementations of three-level qutrits on superconducting hardware, while qudits with $d=7$ states have been successfully realized using trapped $^{40}\mathrm{Ca}^+$ ions \cite{ringbauer_universal_2021}.

The potential benefits of using qudits are especially important in the near-future NISQ (Noisy Intermediate-Scale Quantum) \cite{preskill_quantum_2018} era of quantum computing, as both total qubit count and maximum circuit depth are severely limited by current hardware. Harnessing qudits could allow current or near-future devices to solve larger, more useful problems sooner than with only qubits. However, qudit gate durations theoretically scale in $\mathcal{O}(h^2)$ time \cite{lloyd_efficient_2019}, with $h$ being the Hilbert space dimension; this appears to limit the usefulness of higher-dimension qudits, as experimentally achieving the previously-mentioned decreased ancilla or gate count could require much longer individual gate durations.

\begin{figure}
    \centering
    \includegraphics[width=0.8\linewidth]{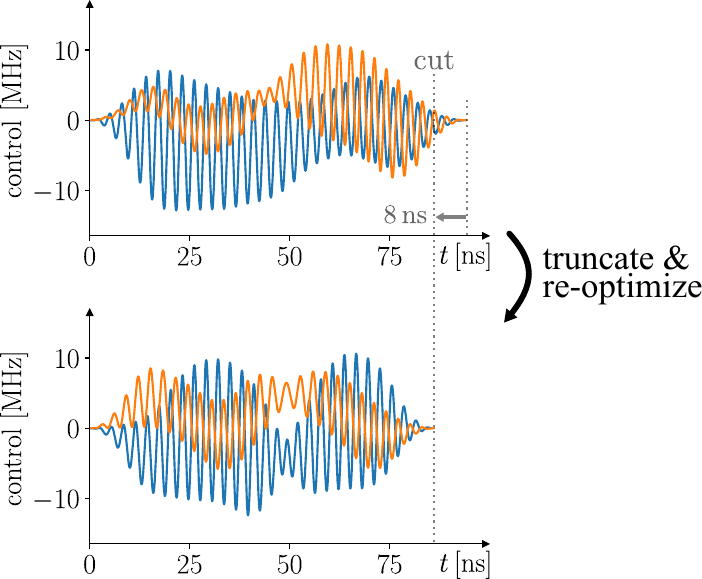}
    \caption{For quantum optimal control tasks to realize a certain gate or state evolution, truncated control pulses can be used as initial guesses for optimizations of shorter duration. Since they originate from a previous optimization, they already drive the system close to the desired objective. In this figure, the plot on the top shows a high-fidelity pulse being truncated to a shorter duration, and the plot on the bottom is the result of re-optimizing this truncated pulse (achieving the same high fidelity at a shorter duration).}
    \label{fig:pulse_cut}
\end{figure}



Quantum logic gates are implemented through external control pulses, meaning that the time it takes to apply a certain gate depends on the duration of the corresponding control pulse. Optimal control software such as GRAPE \cite{khaneja_optimal_2005} is designed to find the most accurate control pulse for a certain gate and duration. This means that finding the high-accuracy pulse of \textit{shortest} duration for a given gate requires many independent optimizations at different durations. In this work, we present the Incremental Pulse Re-seeding (IPR) scheme, a method for repeatedly changing the pulse duration and using previous results as new initial guesses for the optimizer, as shown in the example in Figure \ref{fig:pulse_cut}. IPR provides improved shortest-duration results and decreases the dependence on random initial guesses, which is increasingly important for larger Hilbert space dimension.


Using a transmon Hamiltonian, we apply our method to optimizing single-qudit gates of dimension up to eight and two-qudit gates for qudits of dimension up to four to determine the relationship between Hilbert space dimension and shortest gate duration. We recover the theoretical $\mathcal{O}(h^2)$ duration scaling but find it relatively weak for the range of dimensions we consider, giving a promising near-linear relationship.

In Section \ref{sec:background} we discuss the basics of quantum computation on qubits and qudits. In Section \ref{sec:setup} we explain carrier wave-based optimal control and justify the physical and computational parameters that we choose in our optimizations. In Section \ref{sec:scheme}, we introduce the Incremental Pulse Re-seeding scheme. Finally, we present our optimized pulses and analyze the scaling of gate duration with respect to Hilbert space dimension in Section \ref{sec:results}.

\section{Background} \label{sec:background}
\subsection{Quantum computation with qubits}
In this section, we briefly introduce quantum computation on qubits, and then present the more general approach involving qudits. For a more thorough explanation of quantum computing, we refer the reader to \cite{nielsen_quantum_2000}.

The quantum counterpart to the classical bit is the qubit, the traditional basis of quantum computation. While classical bits can only assume values of either 0 or 1, qubits may exist in a quantum superposition of the corresponding basis states 
\begin{equation}
    \ket{0} = \mqty[1\\0] ~ \text{and} ~ \ket{1} = \mqty[0\\1].
\end{equation}
An arbitrary qubit state is then represented by $\ket{\psi} = \alpha \ket{0} + \beta \ket{1}$,
where $\alpha, \beta \in \mathbb{C}$ denote the complex amplitudes of the respective basis states.

Instructions to manipulate the states of qubits are called quantum gates. According to the laws of quantum theory, these gates are reversible and can be represented by unitary operators. A set of commonly used gates are the Pauli operators:
\begin{equation}
    \label{eq:paulis}
    X = \mqty[0&1\\1&0], \quad Y = \mqty[0&-i\\i&0], \quad Z = \mqty[1&0\\0&-1].
\end{equation}
Here the $X$ gate corresponds to the quantum analogue of a $\mathrm{NOT}$ operation, mapping the state $\ket{\psi} = \alpha \ket{0} + \beta \ket{1}$ to $X\ket{\psi} = \alpha \ket{1} + \beta \ket{0}$. Useful gates for creating superposition or manipulating the relative phases of qubits are the Hadamard gate $H$ and the $T$ gate, respectively:
\begin{equation}
    \label{eq:h2t2}
    H = \frac{1}{\sqrt{2}}\mqty[1&1\\1&-1], \quad T = Z^{\nicefrac{1}{4}} = \mqty[1&0\\0&e^{\nicefrac{i\pi}{4}}].
\end{equation}

Two-qubit gates involve logical operations on several qubits at once, with the most common example being the $\mathrm{CNOT}$ (controlled $\mathrm{NOT}$) gate:
\begin{equation}
    \label{eq:cnot}
    \mathrm{CNOT} \ket{q_1} \ket{q_2} = \ket{q_1} \ket{q_2 \oplus q_1}.
\end{equation}
Like its classical version, this gate flips the state of the target qubit $\ket{q_2}$ if the control qubit $\ket{q_1}$ is in state $\ket{1}$. The $\mathrm{CNOT}$ gate plays a particularly important role in quantum circuits, as it has the power to create entanglement between qubits. Swapping the states of two qubits can be realized with a $\mathrm{SWAP}$ gate:
\begin{equation}
    \label{eq:swap}
    \mathrm{SWAP} \ket{q_1} \ket{q_2} = \ket{q_2} \ket{q_1}.
\end{equation}

The $H$, $T$, and $\mathrm{CNOT}$ gates comprise a universal gate set, meaning any arbitrary qubit circuit can be decomposed into a sequence of these gates.

\subsection{From qubits to qudits}
In many proposed quantum hardware systems, such as superconducting or ion trap computers, each quantum logical unit has an infinite spectrum of energy levels. The standard qubit abstraction suppresses the other states. Instead, we consider $d$-dimensional qu\textit{dits}, which each consist of a superposition of $d$ computational basis states, expressed as
\begin{equation}
    \ket{\psi} = \alpha_0 \ket{0} + \alpha_1 \ket{1} + \dots + \alpha_{d-1} \ket{d-1} = \sum\limits_{k=0}^{d-1} \alpha_k \ket{k}.
\end{equation}

In this work we consider the generalized gates \cite{gottesman_fault-tolerant_1999, yeh_constructing_2022}
\begin{equation}
    \label{eq:dgates}
    \begin{aligned}
        X_d \ket{k} &= \ket{k + 1 \operatorname{mod} d}, \\
        X^s_d: \ket{0} &\leftrightarrow \ket{d-1}, \\
        H_d \ket{k} &= \frac{1}{\sqrt{d}} \sum_{j=0}^{d-1}  \omega^{kj}_d\ket{j}, \\
        T_d \ket{k} &= \omega_d^{\nicefrac{k}{4}} \ket{k},
    \end{aligned}
\end{equation}
where $\omega_d = e^{\nicefrac{2\pi i}{d}}$. $X_d$ maps a computational basis state to the next higher one, and maps $\ket{d-1}$ back to $\ket{0}$. As an alternative generalization, $X^s_d$ swaps the amplitudes of the lowest and the highest state. The generalized Hadamard gate $H_d$ creates an equal-population superposition with different relative phases depending on the input state. In analogy to the qubit case, we define the $T_d$ as the fourth root of the generalized $Z_d$ gate (where $Z_d \ket{k} = \omega_d^k \ket{k}$), such that it preserves populations of the basis states but introduces phase differences. For the above equations, the $d=2$ case produces the qubit gates as defined in \eqref{eq:paulis} and \eqref{eq:h2t2}.

Analogously, two-qudit gates can be generalized as well. For example, the $\mathrm{CNOT}$ operation from \eqref{eq:cnot} can be generalized to the $\mathrm{SUM}$ gate \cite{gottesman_fault-tolerant_1999} or other useful operations depending on use case \cite{gokhale_asymptotic_2019, yeh_constructing_2022}. 
Here we focus on the generalized qudit $\mathrm{SWAP}$ gate, the straightforward extension of \eqref{eq:swap}, which fully swaps the amplitudes of two qudits.

\subsection{Quantum optimal control}
Quantum systems can be purposefully steered by applying external control fields, which are specific to particular hardware. For instance, in superconducting architectures, the state of qubits are manipulated through analog microwave pulses. Generally, the evolution of a quantum system is determined by its time-dependent Hamiltonian
\begin{equation}
    \mathcal{H}(t) = \mathcal{H}_0 + \mathcal{H}_c(t).
\end{equation}
Here $\mathcal{H}_0$ denotes the system's intrinsic drift Hamiltonian and $\mathcal{H}_c(t) = \sum_k f_k(t) \mathcal{H}_k$ denotes the control Hamiltonian, which is typically described by control operators $\mathcal{H}_k$ and classical tunable control fields $f_k(t)$. Quantum optimal control aims to find the optimal control paths $f_k(t)$ to realize a desired state transition or target unitary. This is achieved by repeatedly solving the Schrödinger equation and adjusting the control fields in every iteration to minimize a certain objective function. Different algorithms and toolboxes have been designed for this purpose \cite{khaneja_optimal_2005, petersson_optimal_2021, gunther_quantum_2021}.
\section{Setup} \label{sec:setup}
In this section we motivate the physical and computational parameters that we have chosen to use in our optimizations. We also introduce the concept of control pulses parameterized through B-splines and carrier waves, which is a key aspect of the optimizer we use.

\subsection{Model Hamiltonian}
\label{sub:ham}
We consider a system based on superconducting hardware that consists of two weakly coupled, anharmonic transmons \cite{koch_charge-insensitive_2007} with drift Hamiltonian
\begin{equation}
    \begin{aligned}
        \mathcal{H}_0 = &~\sum_{k=1}^2 \qty[\omega_k a_k^\dagger a_k + \frac{\xi_k}{2} a_k^\dagger a_k^\dagger a_k a_k] \\
        &+ J (a_1^\dagger a_2 + a_2^\dagger a_1).
    \end{aligned}
\end{equation}
In order to describe qudits, we typically truncate the ladder operators $a_k$ and $a_k^\dagger$ at level $d+2$. Including two additional guard states allows us to penalize population leakage into higher energy states, while we have found that additional guard states beyond two provide marginal fidelity improvements for higher computational cost. We choose realistic physical parameters inspired by \cite{sheldon_procedure_2016}: The 0-1 transition frequencies of the transmons are $\omega_1/2\pi = 4.914 \,\mathrm{GHz}$ and $\omega_2/2\pi = 5.114 \,\mathrm{GHz}$, and both transmons have the same anharmonicity $\xi_1/2\pi = \xi_2/2\pi = -0.330 \,\mathrm{GHz}$. They are effectively coupled with $J/2\pi = 3.8 \text{ MHz}$. While our specific results depend explicitly on these parameters, our methods apply to systems with very different parameters. Control of our model system is possible through microwave drives that add or remove single excitations, as described by
\begin{equation}
    \mathcal{H}_c(t) = \sum_{k=1}^2 f_k(t) (a_k + a_k^\dagger).
\end{equation}

To numerically solve the Schrödinger equation, it is often helpful to slow down the time variation by applying the rotating wave approximation. In a rotating frame with equal angular frequency $\omega_r$ for both qudits, we thus obtain the full transformed Hamiltonian
\begin{equation}
    \label{eq:ham_rot}
    \begin{aligned}
        \tilde{\mathcal{H}}(t) = &~ \tilde{\mathcal{H}}_0 + \tilde{\mathcal{H}}_c(t) \\
        = &~\sum_{k=1}^2 \qty[(\omega_k - \omega_r) a_k^\dagger a_k + \frac{\xi_k}{2} a_k^\dagger a_k^\dagger a_k a_k] \\
        &+ J (a_1^\dagger a_2 + a_2^\dagger a_1) \\
        &+ \sum_{k=1}^2 \qty[p_k(t) (a_k + a_k^\dagger) + i q_k(t) (a_k - a_k^\dagger)].
    \end{aligned}
\end{equation}
The rotating frame control functions $p_k(t)$ and $q_k(t)$ are then related to the lab frame controls through $f_k(t) = 2 \Re{(p_k(t) + iq_k(t))e^{i \omega_r t}}$ \cite{petersson_optimal_2021}.

For the case that we want to describe only a single transmon, we disregard the transmon at index $2$ in the above equations and restrict ourselves to a single anharmonic oscillator with one lab frame control field $f_1(t)$.

\subsection{Quantum optimal control with quadratic B-splines}

\begin{figure}[htbp]
    \centering
    \includegraphics[width=\linewidth]{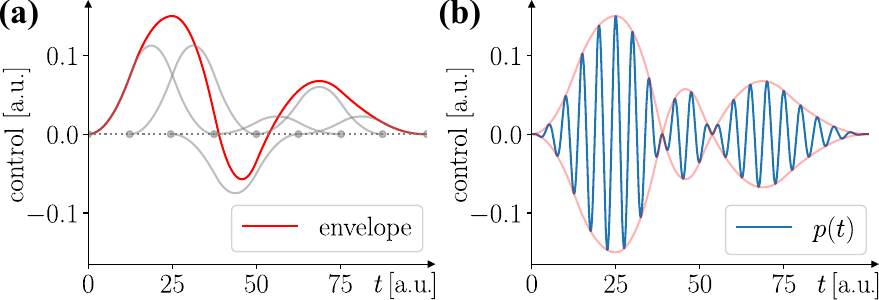}
    \caption{Real part of an exemplary pulse in the rotating frame, with one carrier wave ($N_f=1$). (a) Six B-splines shape the envelope of the pulse. Grey dots mark the support of each time-local basis function. (b) The final pulse is constructed from a single carrier wave bounded by the B-spline envelope. With $N_f>1$, the final pulse is the sum of these individual pulses with one carrier wave.}
    \label{fig:example_pulse}
\end{figure}

Juqbox \cite{petersson_discrete_2020,petersson_optimal_2021} is an open-source software package designed to solve quantum optimal control problems in closed systems under the rotating wave approximation. In this section we briefly summarize the differences between Juqbox and conventional optimizers such as GRAPE \cite{khaneja_optimal_2005, shi_optimized_2019}. While algorithms like GRAPE directly adjust the value of a control pulse at each discrete point in time, Juqbox instead parameterizes the pulse with B-splines. The control functions $p_k(t)$ and $q_k(t)$ in the rotating frame are given by a sum of $N_f$ carrier waves with fixed angular frequencies $\Omega_{k,j}$, where each carrier wave has an amplitude envelope $S_b(t)$ that consists of $N_b$ B-splines:
\begin{equation}
    \label{eq:pq}
    \begin{aligned}
        p_k(t, \vec{\alpha}) &= \sum_{j=1}^{N_f} \sum_{b=1}^{N_b} \Re{\alpha_{k,j,b} \, e^{i \Omega_{k,j} t}} S_b(t) \\
        q_k(t, \vec{\alpha}) &= \sum_{j=1}^{N_f} \sum_{b=1}^{N_b} \Im{\alpha_{k,j,b} \, e^{i \Omega_{k,j} t}} S_b(t). 
    \end{aligned}
\end{equation}
Juqbox finds optimal pulses by adjusting the $2 N_f N_b$ real coefficients $\alpha_{k,j,b} = \alpha_{k,j,b}^{(\mathrm{re})}, + i \alpha_{k,j,b}^{(\mathrm{im})}$ for the quantum devices indexed by $k$. The collection of these design variables is denoted by $\vec{\alpha}$. The benefit of this approach is that the B-spline parameterization drastically reduces the dimensionality of the optimal control problem, especially for long-duration pulses, and allows carrier wave frequencies to be chosen to specifically address the target state evolutions while avoiding undesired transitions. Figure \ref{fig:example_pulse} shows an example pulse defined by a single carrier wave and six B-splines.

The user of this tool must specify a number of physical and numerical parameters for the optimizer, the most significant of which are summarized in Table \ref{tab:optim_params} and discussed in the following subsection.
\begin{table}[htbp]
\caption{Important parameters for optimizing in Juqbox}
\centering
\begin{tabular}{|l|l|}
\hline
\textbf{Symbol} & \textbf{Description} \\
\hline
$T$ & Total pulse duration \\
$\omega_{r,k}$ & Rotating frame frequencies \\
$\tilde{\mathcal{H}}_0$ & Drift Hamiltonian in the rotating frame \\
$\tilde{\mathcal{H}}_c$ & Control Hamiltonian in the rotating frame \\
$\Omega_{k,j}$ & Carrier frequencies in the rotating frame \\
$N_b$ & B-spline basis functions per carrier wave \\
$V$ & Target unitary gate \\
$\alpha_{max}$ & Bound on magnitude of B-splines \\
\texttt{error\_threshold} & Goal infidelity \\
\texttt{max\_iter} & Maximum optimization iterations \\
\hline
\end{tabular}
\label{tab:optim_params}
\end{table}

After specifying the required parameters, Juqbox applies gradient-based methods to minimize an objective function consisting mainly of the trace infidelity 
\begin{equation}
    J = 1 - \frac{1}{h^2} \abs{\Tr{U^\dagger_T(\vec{\alpha})V}}^2
\end{equation}
between the target unitary $V$ and the applied transformation $U_T(\vec{\alpha})$, where $h$ is the Hilbert space dimension (without guard states). The quantum optimal control task is successful if a set of coefficients $\vec{\alpha}$ is found that achieves an infidelity below a desired error threshold. Throughout this work we will also use fidelity $F = 1 - J$ to quantify the quality of pulses. For a given target $V$, certain values of duration $T$ will be too small for the optimizer to find a high-fidelity pulse, in which case a larger $T$ is needed. In our work, the ``gate duration" for a gate $V$ refers to the shortest $T$ for which the optimizer converges to our target fidelity.

\subsection{Choosing optimization parameters}
We observe that for high-dimensional quantum optimal control problems, it is generally difficult to specify a set of optimization parameters that guarantees finding high-fidelity pulses. Parameters such as $N_b$, the number of B-splines per pulse, must be large enough to give the optimizer freedom, while not so large as to introduce unnecessary local minima in the highly non-convex search space. In the following subsection, we describe and justify the decisions we made to choose significant optimization parameters.

The carrier frequency and B-spine envelope parameterization of the control pulse has two benefits: It reduces the dimensionality of the pulse, and additionally allows us to easily target the desired state transitions in the qudits by using carrier frequencies directly corresponding to the qudit states. In the lab frame, the $N_f$ carrier frequencies
\begin{equation}
    \Omega^{\mathrm{lab}}_{k,j} = \omega_k + j \xi_k, \quad j=0,\dots,d-2
\end{equation}
correspond to the resonant frequencies of qudit $k$. For instance, $\Omega_{1,2}^{\text{lab}}$ is the frequency of the $\ket{2} \to \ket{3}$ transition for qudit $1$. For a two-qudit gate, each control pulse includes the complete set of $2(d-1)$ carrier frequencies in order to make use of the cross-resonance effect \cite{rigetti_fully_2010, ware_cross-resonance_2019}, where driving one transmon at a resonant frequency of the other transmon triggers a state transition in the latter. We choose not to correct these frequencies to account for the weak coupling between the transmons, since the small corrections fall inside the widths of the respective Fourier peaks of pulses with finite duration. From experience, we have found that considering all $(d-1)^2$ true resonant frequencies significantly increases the complexity of the problem without providing a noticeable improvement.

Given the fixed set of $2(d-1)$ lab frame carrier frequencies $\Omega^{\mathrm{lab}}_{k,j}$, we choose the average frequency as the rotating frame frequency $\omega_r = (\Omega^{\mathrm{lab}}_{\mathrm{max}} + \Omega^{\mathrm{lab}}_{\mathrm{min}})/2$, which is equal for both qudits (the coupling term is then constant in time, as shown in \eqref{eq:ham_rot}). This minimizes the magnitude of the carrier frequencies in the rotating frame, allowing the optimizer to choose larger discrete time steps and decreasing overall optimization run time. As an example we consider two qutrits ($d=3$). The lab frame carrier frequencies are $\Omega^{\mathrm{lab}}_{k,j} / 2\pi \in \qty{4.914, 4.584, 5.114, 4.784} \,\mathrm{GHz}$, where the first two correspond to the two resonant frequencies of transmon 1 and the latter two to the resonant frequencies of transmon 2. The rotating frame frequency is $\omega_r/2\pi = (5.114 + 4.584)/2 \,\mathrm{GHz} = 4.849 \,\mathrm{GHz}$. Thus, the carrier frequencies in the rotating frame are given by $\Omega_{k,j} / 2\pi = (\Omega^{\mathrm{lab}}_{k,j} - \omega_r)/2\pi \in \qty{0.065, -0.265, 0.265, -0.065} \,\mathrm{GHz}$.

The number of B-spline basis functions $N_b$ determines how fast the pulse envelope can vary over time (see Figure \ref{fig:example_pulse}). While a larger number may enable shorter-duration pulses, this will increase the dimensionality of the problem, as well as introduce non-carrier-frequency control pulse oscillations. We choose $N_b = \qty[T/(10\,\mathrm{ns})] + 2$ (where $\qty[\, . \,]$ denotes the nearest-integer function) for two reasons. First, we want to maintain an approximately constant B-spline density when varying $T$ for consistency between different optimization problems. Second, this choice guarantees a minimum envelope rise time of around 15 ns, which is realistic in experiment \cite{sheldon_procedure_2016}. The term $+2$ accounts for B-splines on the time domain boundaries, which are always set to amplitude $0$ in Juqbox's implementation to ensure that the final pulse starts and ends at $0$.

The parameter $\alpha_{\mathrm{max}}$ limits the amplitudes of the individual B-splines, effectively limiting the maximum amplitude of the final pulses. This is needed to ensure weak driving \cite{magesan_effective_2020}, which increases the accuracy of approximate models like \eqref{eq:ham_rot}. Additionally, limited power reduces the risk of leakage into guard levels due to off-resonant transitions. We aim for lab frame pulse amplitudes of at most $40\,\mathrm{MHz}$ and achieve this by tuning $\alpha_\mathrm{max}/2\pi = 40/(2\sqrt{2} N_f) \,\mathrm{MHz}$.

We set $\texttt{error\_threshold}$, the target gate infidelity, to $10^{-3}$, corresponding to a fidelity of $99.9\%$. The optimization terminates if the pulse fidelity reaches this target.

Finally, the parameter $\texttt{max\_iter}$ limits the number of iterations of the optimizer if it does not terminate early due to reaching $\texttt{error\_threshold}$. This variable must be set sufficiently large to ensure that the optimizer has converged very nearly to a local minimum, and is generally set experimentally depending on the dimension and target unitary (typically in the range $200\dots1000$).

\section{Gate duration optimization} \label{sec:scheme}

\begin{figure*}[htbp]
    \centering
    \includegraphics[width=0.59\linewidth]{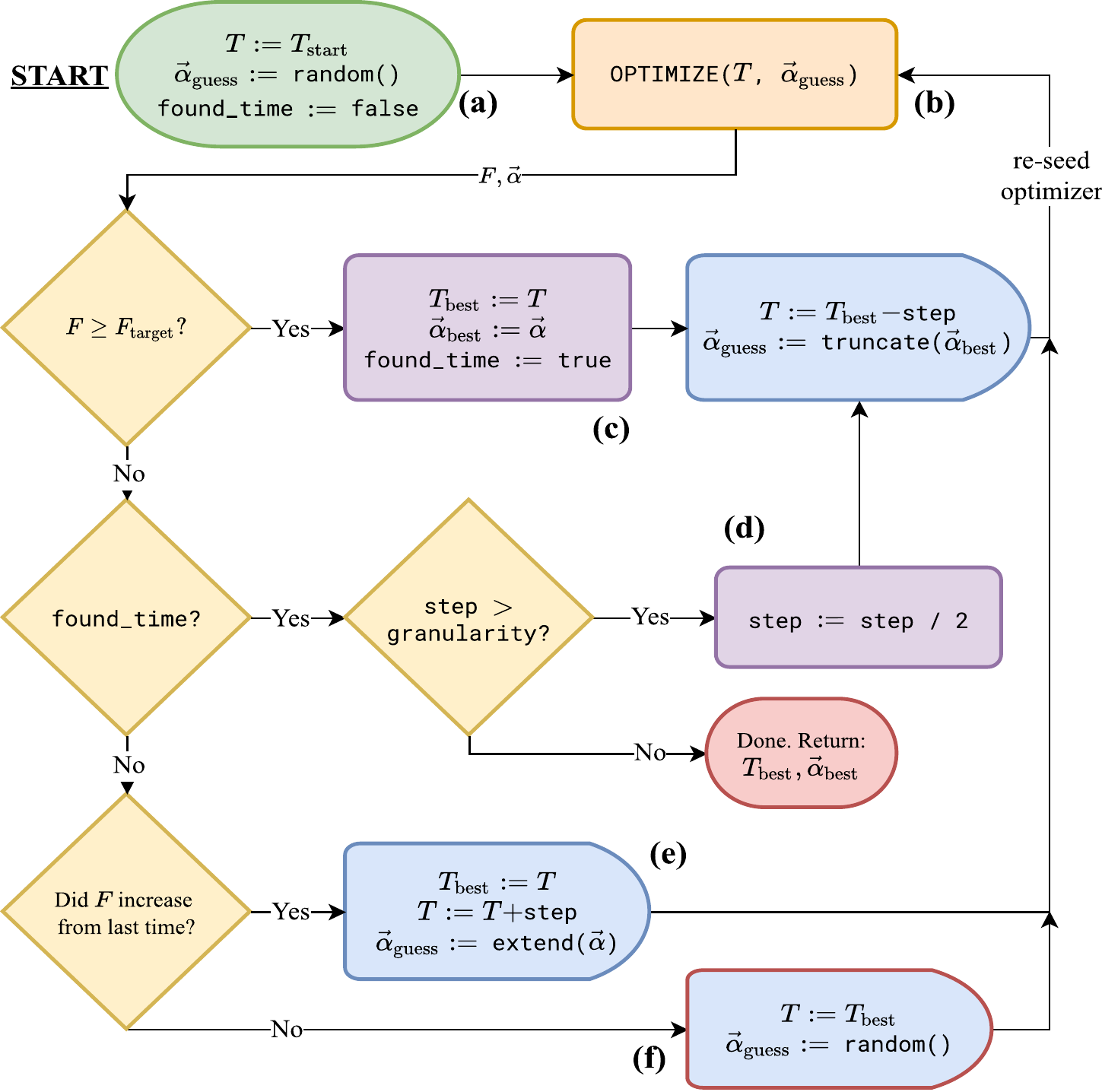}
    \caption{The Incremental Pulse Re-seeding process. (a) The problem is initialized with a random pulse and duration guess $T_\mathrm{start}$. (b) The main optimization function, which searches for the highest-fidelity pulse for a given duration using Juqbox, returning an optimized pulse $\vec\alpha$ and a fidelity $F$. (c) If an optimization succeeds, save the duration and pulse, then decrease $T$ and start from a truncated version of the successful pulse. (d) If an optimization fails but a successful pulse has been found in the past, decrease step size and work backwards from $T_\mathrm{best}$ again; if step size has reached granularity, return best pulse. (e) If no successful pulse has been found so far, but fidelity improved, mark this duration as the best so far, extend the duration, and re-seed with the extended previous pulse. (f) If no successful pulse has been found and fidelity decreased, restart the whole process with the highest-fidelity duration and a random guess.}
    \label{fig:flowchart}
\end{figure*}

\begin{figure*}[htbp]
    \centering
    \includegraphics[width=0.80\linewidth]{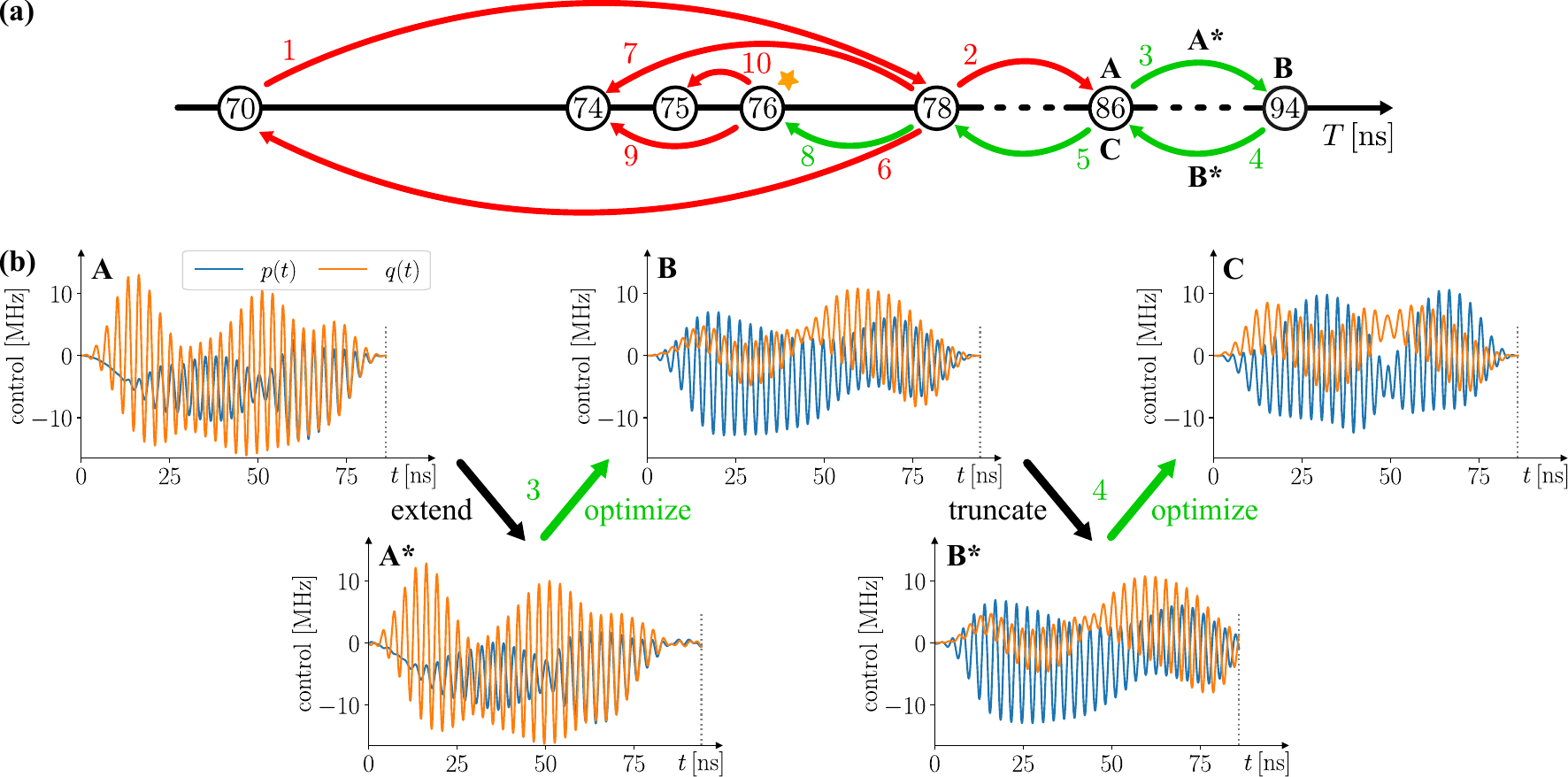}
    \caption{Applying Incremental Pulse Re-seeding (IPR) to find the shortest duration for the $H_4$ gate. (a) Visualization of attempted durations during the IPR procedure, with the chronological order of steps indicated by the colored numbers. Green (red) arrows represent steps that did (did not) lead to meeting the pulse target fidelity of $99.9\%$. Starting from $T_\mathrm{start} = 70\,\mathrm{ns}$ and a time step of 8\,ns, three pulse extension and re-optimization steps are required before the target fidelity is first reached at $94\,\mathrm{ns}$. The duration is then reduced to $78\,\mathrm{ns}$ by successfully re-seeding with truncated pulses. After step $6$ the time step is halved because re-optimizing at $T=70\,\mathrm{ns}$ with the truncated pulse does not meet the target fidelity. Repeating this scheme leads to the final duration of $T_\mathrm{gate} = 76\,\mathrm{ns}$, highlighted by the star, as after step $10$ the time step falls below the granularity of $1\,\mathrm{ns}$. (b) Plot A shows the optimized pulse after IPR step $2$. A new seed for the optimization at $T=94\,\mathrm{ns}$ is generated by extending the previous result in duration (A*). At $94\,\mathrm{ns}$ the target fidelity is met, so the optimized pulse is truncated (B*) and used as a seed for the second $T=84\,\mathrm{ns}$ optimization, leading to the successful result shown in C.}
    \label{fig:H4_IPR}
\end{figure*}

In this section we describe our method for finding the shortest-duration, high-fidelity pulse for a given target unitary. Optimal control software generally works with a fixed pulse duration, so finding the shortest high-fidelity pulse involves repeated optimizations at different durations.

By default, each optimization begins with a random guess for the coefficient vector $\vec\alpha$ within some bounds for each coefficient (typically $\pm\alpha_\mathrm{max}/100$ or $\pm\alpha_\mathrm{max}/10$). Due to the non-convex nature of the optimization space, it is generally good practice to try multiple initial guesses if the first does not work and the fidelity seems relatively close to the goal fidelity.

\subsection{Naive approaches to duration optimization}

We first introduce two simple approaches to finding optimal gate durations, the exhaustive method and the binary search method, and explain their downsides. The exhaustive process begins at some low $T=T_\mathrm{start}$ and increases $T$ at a fixed interval, running an independent random-guess optimization for each duration, until one value of $T$ yields a high-fidelity pulse. While this method will work (at least for low-dimension qudits), it is computationally inefficient, motivating the second approach: a binary search. This method involves specifying $T_\mathrm{min}$ and $T_\mathrm{max}$ and then optimizing in a binary search pattern within the bounds, decreasing $T$ if optimization succeeds and increasing if it fails (still running an independent, random-guess optimization at each duration) until reaching the specified granularity.

While these approaches are relatively effective for small qubit-based gates, they both have problems when transitioning to high-dimensional qudit gates. Larger qudit dimension significantly increases computational cost of solving the optimization, which means the exhaustive method can potentially be very computationally expensive if the starting $T$ is far below the shortest convergent duration. The binary search method is more efficient, but individual long-duration optimizations near $T_\mathrm{max}$ will still be quite expensive if $T_\mathrm{max}$ is large.

However, the more significant issue is the increased complexity of the optimization space for higher qudit dimensions. We find that random guesses become less effective at consistently converging to the optimal solution for a given $T$, as we show in Section \ref{ss:random}.

\subsection{Incremental Pulse Re-seeding}
We have developed an optimization scheme to address both of these problems. The key idea is to reuse the coefficient vector $\vec\alpha$ from failed optimizations as a seed for the next. For example, if a pulse of duration $T_0$ converges to a fidelity of $99.5\%$, we can reuse this pulse as the starting point for an optimization at duration $T_0 + \delta$, which we find to have a better chance of converging to $99.9\%$ fidelity than a random guess.

Depending on the success of the optimizer to find a high-fidelity pulse for a certain duration, our algorithm changes the duration for the next iteration by a discrete time step and reuses the previous pulse (by extending or truncating it to fit the new duration) as a seed. The step size decreases as the algorithm approaches the final solution. We therefore call this scheme \emph{Incremental Pulse Re-seeding} (IPR). The variables used in IPR are listed in Table \ref{tab:scheme_params}, and a flowchart of the method is shown in Figure \ref{fig:flowchart}. We denote an optimized pulse as ``successful" if it has fidelity greater than our goal fidelity, and ``failed" if not.

\begin{table}[htbp]
\caption{Variables used in Incremental Pulse Re-seeding}
\centering
\begin{tabular}{|l|l|}
\hline
\textbf{Symbol} & \textbf{Description} \\
\hline
$T_\mathrm{start}$ & First duration tried by optimizer \\
\texttt{step} & Time step to increase/decrease by \\
\texttt{granularity} & Minimum step size \\
$\vec\alpha_\mathrm{guess}$ & B-spline coefficients of initial guess pulse\\
\texttt{error\_threshold} & Target infidelity \\
\hline
\end{tabular}
\label{tab:scheme_params}
\end{table}

The program starts from an initial duration $T_\mathrm{start}$ with a random guess $\vec\alpha_\mathrm{guess}$. A flag variable \texttt{found\_time}, which indicates if a successful pulse has been found before, is set to \texttt{false}. The optimizer then converges to a solution $\vec\alpha$ with fidelity $F$. If the converged pulse is of sufficiently high fidelity, the solution is stored as $\vec\alpha_\mathrm{best}$, the duration is stored as $T_\mathrm{best}$, and \texttt{found\_time} is set to \texttt{true}. We then reduce the gate duration by the step size and re-seed the optimizer with a truncated version of $\vec\alpha_\mathrm{best}$ that fits the new duration. Figure $\ref{fig:pulse_cut}$ shows an example of a pulse being truncated and re-optimized.

If the target fidelity is not met, but a successful pulse has been found before, we decrease the step size and repeat the above steps, starting from $T_\mathrm{best} - \texttt{step}$ with $\vec\alpha_\mathrm{guess}=\texttt{truncate}( \vec\alpha_\mathrm{best} )$. This only happens if the step size is larger than the predefined granularity limit (in our case $1\,\mathrm{ns}$); otherwise, the algorithm terminates and returns the converged result with the shortest duration.

If the optimized pulse does not meet the target fidelity, and no successful pulse has been found before, we increase the duration and re-seed the optimizer with an extended version of the previous pulse, for as long as the fidelity increases with each optimization. If the fidelity decreases (compared to a shorter duration) before any solution is found, we start the entire procedure with a new random $\vec\alpha_\mathrm{guess}$. In this case we set $T_\mathrm{start}=T_\mathrm{best}$, the duration for which fidelity was highest. 

\subsection{Example: $H_4$ gate}
As an example, we present an application of Incremental Pulse Re-seeding to find the optimal duration of the $d=4$ Hadamard gate $H_4$ (see \eqref{eq:dgates}). Figure \ref{fig:H4_IPR}(a) shows the sequence of IPR steps, where the chronological order is given by the numbered arrows. Red arrows represent steps that failed to reach the target fidelity of $99.9\%$, while green arrows indicate success. We start at $T_\mathrm{start}=70\,\mathrm{ns}$ with an initial time step size of $8\,\mathrm{ns}$ and a random guess for the coefficient vector $\vec\alpha$. The fidelity of the optimized pulse falls short of the target, so IPR extends the output pulse and uses it to seed the optimizer at a longer duration (step $1$). This occurs two more times (steps $2$ and $3$) before a pulse that reaches the target fidelity is found at $T=94\,\mathrm{ns}$. In step $4$, this solution is then truncated and used to revisit $T=86\,\mathrm{ns}$. This time the optimization task is successful, so the duration can be further decreased (step $5$), resulting in another high-fidelity pulse at $78\,\mathrm{ns}$. $T=70\,\mathrm{ns}$ fails to converge to a the target fidelity, so the time step is halved and IPR tries $74\,\mathrm{ns}$ instead (step $7$). After few more steps, the minimum step granularity is reached, and $T_\mathrm{gate}=76\,\mathrm{ns}$ is found to be the shortest duration to realize the $H_4$ gate with $99.9\%$ fidelity. It is important to note that durations $78\,\mathrm{ns}$ and $86\,\mathrm{ns}$ both initially resulted in failed optimizations, but then succeeded when re-seeded with truncated successful pulses.

\begin{figure}[htpb]
    \centering
    \includegraphics[width=\linewidth]{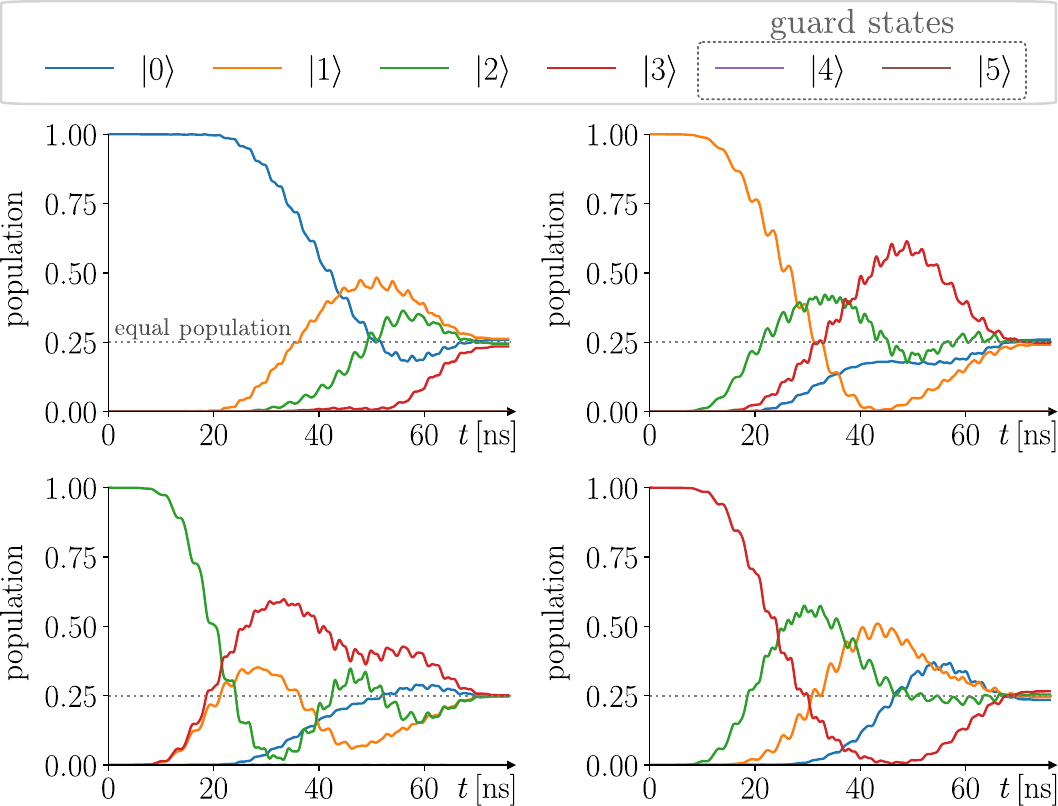}
    \caption{Evolution of state populations when applying the $H_4$ pulse of shortest duration ($76\,\mathrm{ns}$) to a qudit with $d=4$. Every computational basis state is transformed into superposition of all four basis states with nearly equal population, deviations only arising from the remaining $0.1\%$ infidelity. At all times the populations of the guard states $\ket{4}$ and $\ket{5}$ are below $2 \times 10^{-3}$ and $10^{-6}$, respectively.}
    \label{fig:H4_states}
\end{figure}

With the help of Figure \ref{fig:H4_IPR}(b), we take a closer look at steps $3$ and $4$ to explain how pulses are extended and truncated. Initially (plot A), we have an optimized $86\,\mathrm{ns}$ pulse that falls short of the target fidelity. The controls $p(t)$ and $q(t)$ are numerically extended by appending $8\,\mathrm{ns}$ of idle time (zero amplitude pulse). The number of B-spines $N_b$ changes due to the duration change, so we obtain the new coefficient vector $\vec\alpha_\mathrm{guess}$ by applying a least-squares fit according to \eqref{eq:pq} with the new number of B-splines. The approximation introduces small distortions, as can be seen in plot A*, because the manually modified control functions are not perfectly realizable with the given parameterization. The following optimization (plot B) meets the target fidelity, so the pulse is truncated to $86\,\mathrm{ns}$ and re-parameterized (plot B*). Plot C shows the converged pulse after step $4$, which successfully reaches $99.9\%$ fidelity. Together, the series of plots (where A and C are both at duration $86\,\mathrm{ns}$) shows how a new minimum was found after visiting $94\,\mathrm{ns}$ and then revisiting $86\,\mathrm{ns}$ with a better guess.

Figure \ref{fig:H4_states} visualizes the evolution of the basis state populations for the shortest pulse of $76\,\mathrm{ns}$. It is clearly visible how an equal-population superposition emerges in every case, and the small imperfections at the final time come from the remaining $0.1\%$ infidelity. The populations of the guard states cannot be seen because they are suppressed by multiple orders of magnitude. This guard state suppression is achieved by the carrier wave parameterization method, which allows us to selectively induce state transitions.

\subsection{Comparison with random guessing} \label{ss:random}
\begin{figure}[htbp]
    \centering
    \includegraphics[width=\linewidth]{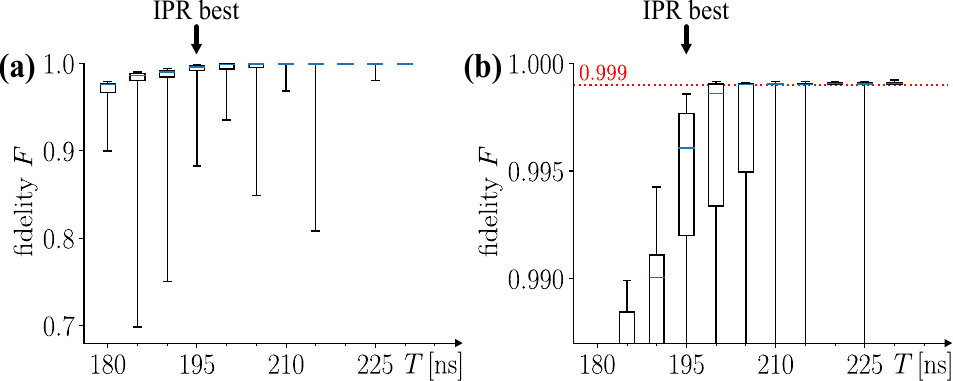}
    \caption{Distribution of naive random-guess optimizations of the $X_8$ gate for invididual durations between $180\,\mathrm{ns}$ and $230\,\mathrm{ns}$. Each boxplot consists of $20$ fidelities from $20$ individual Juqbox optimizations, with the blue line indicating the median value, the box indicating the middle $50\%$ of fidelities, and the whiskers indicating the full range of values. (a) The full data, showing a general increase in fidelity as duration increases. (b) Focusing on the area of interest (fidelities close to $99.9\%$). The IPR scheme outperforms random guessing, finding a minimum duration of $T_\mathrm{best}=195\,\mathrm{ns}$ at $99.9\%$ fidelity, whereas none of the 20 random guesses at that duration reached the same fidelity.}
    \label{fig:naive}
\end{figure}

For the $X_8$ gate (see \eqref{eq:dgates}), the IPR scheme finds a shortest-duration pulse of $195\,\textrm{ns}$ with target fidelity $99.9\%$. To compare our approach with a naive approach, we run 20 random-guess, single-duration Juqbox optimizations at every $5\,\mathrm{ns}$ between $180\,\mathrm{ns}$ and $230\,\mathrm{ns}$, where the components of $\vec\alpha_\mathrm{guess}$ are sampled from $[-0.1\alpha_\mathrm{max}, 0.1\alpha_\mathrm{max}]$. Figure \ref{fig:naive} shows the distribution of pulse fidelities achieved by this naive random guessing across a range of durations for the same gate, with 20 random attempts at each duration.

It is evident that there is significant variation in the naive approach; no random-guess optimizations converged to $99.9$ fidelity at $195\,\mathrm{ns}$, and only $8$ out of $20$ random guesses converged to $99.9\%$ fidelity at $200\,\mathrm{ns}$. Even for larger durations such as $215\,\mathrm{ns}$, not all guesses converged to the target fidelity. This is clearly a problem for binary-search-based methods, which implicitly assume that a given duration will either always succeed or never succeed. In contrast, over ten IPR runs (with $T_\mathrm{start}$ sampled from $[160,240]\,\mathrm{ns}$; see Section \ref{ss:single}), we find an average duration of $198.6\,\mathrm{ns}$ and a standard deviation of $3.0\,\mathrm{ns}$. This small standard deviation given the large range of starting times is evidence that our method is more stable than random guessing for large Hilbert space optimizations.

\section{Qudit pulse results} \label{sec:results}

\begin{figure*}[htbp]
    \centering
    \includegraphics[width=\linewidth]{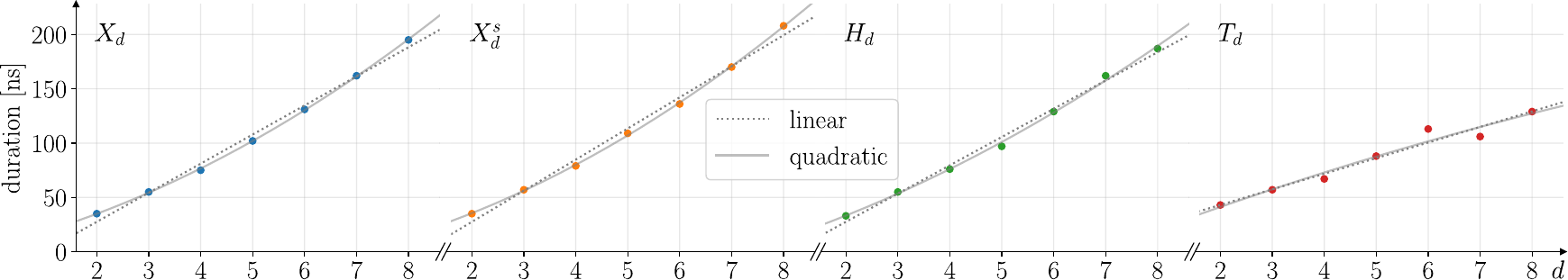}
    \caption{Dependence of single-qudit gate durations from qudit dimension $d$. The four gates are defined in \eqref{eq:dgates}. Each data point shows the minimum duration of applying Incremental Pulse Re-seeding (IPR) from ten different starting times. The unexpected discrepancy between $d=6$ and $d=7$ for the generalized $T_d$ gate suggests a highly complicated optimization space and shows that shorter pulses for some $d$ values may be possible. For each gate the scaling of duration with dimension $d$ is analyzed with quadratic and linear regressions, where the latter is less accurate but captures the essential behavior in this low-dimensional regime. The fit coefficients are presented in Table \ref{tab:fit_coeffs}.}
    \label{fig:onequdit}
\end{figure*}

Previous work \cite{gokhale_asymptotic_2019}, \cite{baker_efficient_2020} has shown that breaking the qubit abstraction in quantum circuits can lead to substantially reduced circuit complexity involving fewer ancilla, making qudits promising candidates for quantum computation. However, these improvements can only be practical if the gate duration overhead does not scale poorly with the Hilbert space dimension while still maintaining high gate fidelity. In this section we investigate this relation by explicitly constructing short duration pulses of at least $99.9\%$ fidelity for one- and two-qudit gates using Incremental Pulse Re-seeding (IPR).

\subsection{Single-qudit gates} \label{ss:single}

We consider the single-qudit gates $X_d$, $X^s_d$, $H_d$ and $T_d$ presented in \eqref{eq:dgates} for dimensions $d=2,...,8$. In each case we include two guard levels to account for and suppress leakage into higher energy states. Expecting short gate durations, we choose $T_\mathrm{start} = 50\,\mathrm{ns}$ for the qubit and qutrit ($d=3$) cases, and use the resulting shortest durations to estimate $T_\mathrm{start}$ for higher $d$ with a linear extrapolation. We set $\texttt{step}$ to the power of two nearest to $0.1 \, T_\mathrm{start}$ because this results in integer durations. Having found durations $T_\mathrm{best}$ for every dimension, we then run IPR repeatedly with starting durations randomly chosen from $[0.8 \, T_\mathrm{best}, 1.2 \, T_\mathrm{best}]$ to identify other local minima and reduce the gate time further, as well as to measure the uncertainty of the method.

Figure \ref{fig:onequdit} shows the minimum single-qudit gate durations found over ten IPR runs for each dimension. The durations for the generalized gates $X_d$, $X_d^s$ and the Hadamard gate, $H_d$, scale similarly with the qudit dimension. We suspect that the generalized $T_d$ gate can be realized in lower time at higher dimension compared to the others because it only causes a change in phase, not in state populations.
For this gate, we observe that the best pulse found for $d=7$ is of shorter duration than the $d=6$ pulse, when the reverse is intuitively expected. This discrepancy suggests shorter pulses may exist and indicates a highly complicated optimization space.
Theoretical studies \cite{lloyd_efficient_2019} have shown that optimal gate durations scale asymptotically with $\mathcal{O}(h^2)$, where $h$ is the dimension of the Hilbert space. We therefore fit each sequence of data points with a quadratic polynomial $T_\mathrm{quad}(d) = ad^2 + bd + c$ (solid grey lines) in $d=h$ to compare our findings with the theory. Furthermore, the layout of the data also motivates linear regression, so we investigate the performance of a linear model $T_\mathrm{lin}(d) = bd + c$ (dotted grey lines) as well. The fit coefficients and their standard deviation estimates are given in Table \ref{tab:fit_coeffs}, where for each gate the first line corresponds to the linear fit and the second line to the quadratic fit. The quality of each regression is quantified with the coefficient of determination $R^2$. 

\begin{table}[htbp]
\caption{Linear and quadratic fit coefficients for single-qudit gate durations}
\centering
\begin{tabular}{|c|c|c|c|c|}
\hline
\textbf{Gate} & $a$\,[ns] & $b$\,[ns] & $c$\,[ns] & $R^2$ \\
\hline
\multirow{2}{*}{$X_d$} & & $26.79 \pm 1.16$ & $-26.07 \pm 6.23$ & $0.991$ \\
& $1.48 \pm 0.12$ & $12.02 \pm 1.18$ & $4.93 \pm 2.67$ & $1.000$ \\[2pt]
\multirow{2}{*}{$X^s_d$} & & $28.64 \pm 1.26$ & $-29.79 \pm 6.78$ & $0.990$ \\
& $1.60 \pm 0.16$ & $12.69 \pm 1.60$ & $3.71 \pm 3.63$ & $1.000$ \\[2pt]
\multirow{2}{*}{$H_d$} & & $26.04 \pm 1.05$ & $-24.61 \pm 5.68$ & $0.992$ \\
& $1.15 \pm 0.36$ & $14.49 \pm 3.66$ & $-0.36 \pm 8.28$ & $0.998$ \\[2pt]
\multirow{2}{*}{$T_d$} & & $14.36 \pm 1.37$ & $14.36 \pm 7.36$ & $0.957$ \\ & $-0.38 \pm 0.86$ & $18.17 \pm 8.74$ & $6.36 \pm 19.78$ & $0.959$ \\
\hline
\end{tabular}
\label{tab:fit_coeffs}
\end{table}
The near-ideal $R^2$ values for quadratic fits to $X_d$, $X^s_d$ and $H_d$ indicate that the quadratic models describe the empirical scaling behaviors for these gates very well, which agrees with the theoretical predictions. However, we emphasize that linear scaling models capture the essential behavior in this low-dimensional regime (compared to asymptotic considerations) in good approximation too, deviating from the duration points by less than $5\,\mathrm{ns}$ on average for those three gates. Barely any difference between quadratic and linear regression is noticeable for the $T_d$ gate, but the more scattered data points lead to a reduced $R^2$ value and larger uncertainty. Additionally, the fit parameter $a < 0$ indicates an unrealistic scaling of pulse duration because this leads to negative durations in the limit of large $d$. 
$T_d$ is a good case study in the interplay between theoretical expectations and empirical realizations. Specifically, the poor fit obtained here indicates that for many values of $d$, these gate durations may overestimate the optimal gate times.

This result of essentially linear scaling over the practical range of qudit dimensions has promising implications for the current era where qudit experiments have been conducted successfully \cite{galda_implementing_2021, rigetti_computing_beyond_2021, ringbauer_universal_2021}. It suggests that the computational advantage single-qudit gates can provide is not outweighed by pulse time overhead.


\begin{figure}[htbp]
    \centering
    \includegraphics[width=0.62\linewidth]{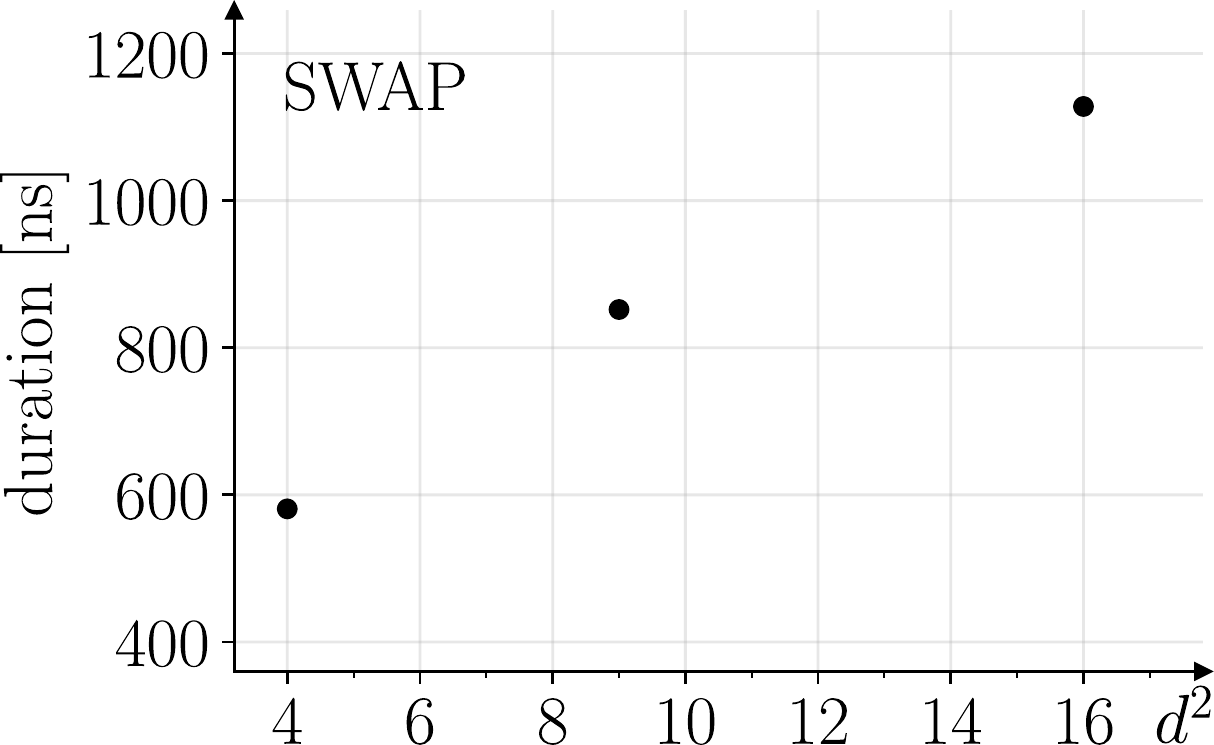}
    \caption{Gate durations of two-qudit $\mathrm{SWAP}$ operations for qudit dimensions $d=2, 3, 4$. The horizontal axis corresponds to the dimension of the full Hilbert space, $h=d^2$. Due to the low number of data points, regression is not applied, but a near-linear trend is visible.}
    \label{fig:twoqudit}
\end{figure}

\subsection{Two-qudit $\textrm{SWAP}$ gate}

For the two-qudit case we consider the generalized $\mathrm{SWAP}$ gate, which fully swaps the state amplitudes between the qudits. In many architectures, this is an important operation for communication, as limited connectivity between devices requires quantum information to be moved around repeatedly. We apply IPR to find $\mathrm{SWAP}$ durations for two qudits of dimensions $2$, $3$, and $4$, with corresponding Hilbert space dimensions $4$, $9$, and $16$, respectively. We limit our analysis to these three cases because simulating the evolution of two qudits with $d \geq 5$ (plus guard states) becomes very resource-expensive. We use two guard states for dimensions $2$ and $3$ and only one guard state for $4$ due to the large Hilbert space. As in the single-qudit case, each data point is the minimum duration from ten IPR runs with different $T_\mathrm{start}$ values.

Figure \ref{fig:twoqudit} shows shortest-duration results for the $\mathrm{SWAP}$ gate with respect to the Hilbert space dimension $h=d^2$. 
Given the small number of points, we do not attempt a fit, but note that the points appear nearly linear, as was the case with the single-qudit gates. It is worth noting that fully swapping the states of two ququarts ($d=4$) is slightly less expensive ($1128\,\mathrm{ns}$) than performing two sequential $\mathrm{SWAP}$ operations on qubits ($581\,\mathrm{ns}+581\,\mathrm{ns}=1162\,\mathrm{ns}$). Given that a single ququart has the same information capacity as two qubits, this represents an efficient way of moving quantum information. Additionally, given that the $d=4$ $\mathrm{SWAP}$ optimization space is much more complex than the qubit case, we expect that this duration could be further decreased through more sampling or fine-tuning.
\section{Discussion} \label{sec:discussion}

We have presented the Incremental Pulse Re-seeding (IPR) scheme, an efficient method for finding short-duration pulses for quantum optimal control tasks, which performs better than random-guess based methods for large Hilbert space gates. We use this method together with the optimal control software Juqbox to demonstrate that single-qudit gate duration scaling is nearly linear for several useful gates, with dimension up to $d=8$. We also show that the two-qudit $\mathrm{SWAP}$ gate appears to follow a similar trend for Hilbert space dimensions up to $d^2=16$.

We emphasize that while Incremental Pulse Re-seeding helps to make intelligent initial pulse guesses for the optimizer during intermediate steps, it does not eliminate the possibility of converging to a local infidelity minimum. As quantum optimal control problems are typically underconstrained, different initial guesses can converge to different solutions. In our case, this means starting IPR with different values for $T_\mathrm{start}$ and $\vec\alpha_\mathrm{guess}$ may result in slightly different gate durations. However, as an example, our simulations show a relatively small variation in gate duration of $4.69\,\mathrm{ns}$ on average for single-qudit gates, underlining the advantage of our algorithm.

Our results prompt several directions of future research:
\begin{itemize}
    \item The efficiency of qudit gates motivates further exploration into quantum circuits that explicitly use qudits for computation, such as the asymptotic ancilla reductions achieved in \cite{gokhale_asymptotic_2019} and \cite{baker_efficient_2020}. The computational advantages of qudit circuits may allow NISQ devices to solve previously intractable problems.
    \item In this work, we consider the generalized $\mathrm{SWAP}$ two-qudit gate, which appears to scale nearly linearly for small qudit dimension. This result encourages analysis of duration scaling for more two-qudit gates, such as the $\mathrm{SUM}$ gate (a generalization of $\mathrm{CNOT}$) from \cite{gottesman_fault-tolerant_1999}.
    \item The presented results are based on a system Hamiltonian modeling superconducting transmons. We emphasize that our approach is not limited to this specific case and could be directly applied to optimal control problems across a wider class of quantum systems.
    \item Our optimizations use an approximate Hamiltonian in a closed system (governed by the Schrodinger equation). For a more accurate measure of gate duration scaling, similar experiments are needed in an open system, where errors can be modeled more realistically. This can be achieved by describing the dynamics with the GKSL master equation \cite{lindblad_generators_1976, gorini_completely_1976}. In addition, benchmarking the performance of qudit pulses on actual noisy quantum architectures would help to investigate the accuracy of models employed. Optimizing control pulses to specifically mitigate the increased sources of errors from operating on higher energy levels may lead to making qudit computation practical.
\end{itemize}

Overall, our work highlights the potential of qudit-based computation in the future and provides an effective method for finding short-duration qudit pulses. We find high-fidelity pulses of low duration for both one- and two-qudit gates with nearly linear scaling in the hardware-practical regime, suggesting that qudit computation can offer significantly increased efficiency compared to qubit-only circuits.
\section*{Acknowledgements}
We would like to thank Stefanie Günther and N. Anders Petersson for valuable advice on using the quantum optimal control software packages Juqbox and Quandary. Additionally, we would like to thank David I. Schuster for helpful discussions regarding quantum optimal control theory.

We are grateful for the support of the University of Chicago’s Research Computing Center for assistance with the calculations carried out in this work.

\bibliography{bibliography.bib}

\end{document}